\documentclass{IEEEtran}
\usepackage{newtxtext}

\usepackage{cite}
\usepackage{amsmath,amssymb,amsfonts}
\usepackage{algorithmic}
\usepackage{graphicx}
\usepackage{textcomp}

\def\BibTeX{{\rm B\kern-.05em{\sc i\kern-.025em b}\kern-.08em
    T\kern-.1667em\lower.7ex\hbox{E}\kern-.125emX}}

\usepackage{tabularx}
\usepackage{amsfonts}
\usepackage{booktabs}
\usepackage{newtxtext,newtxmath}
\usepackage{bm}

\usepackage{multirow}
\usepackage{float}
\usepackage{algorithm}
\usepackage{scalerel}

\usepackage{makecell}
\usepackage[hidelinks]{hyperref}

\begin{document}

\title{Optimized Sampling of Angle-Resolved Scatterometry Data Using End-to-End Compressed Learning Model for Nanograss Deficiency Detection}

\author{
Mehdi Abdollahpour,~\IEEEmembership{Student Member,~IEEE,}
Carsten Bockelmann,~\IEEEmembership{Member,~IEEE,}
and Armin Dekorsy,~\IEEEmembership{Senior Member,~IEEE}
\thanks{Mehdi Abdollahpour, Carsten Bockelmann, and Armin Dekorsy are with the Department of Communications Engineering, University of Bremen, 28359 Bremen, Germany.}
\thanks{Corresponding author: Mehdi Abdollahpour, e-mail: abdollahpour@ant.uni-bremen.de.}
}

\markboth
{Abdollahpour \MakeLowercase{\textit{et al.}}: Optimized Scatterometry Sampling for Nanograss Deficiency Detection}
{Abdollahpour \MakeLowercase{\textit{et al.}}: Optimized Scatterometry Sampling for Nanograss Deficiency Detection}

\maketitle

\thispagestyle{empty}

\begin{abstract}
Reliable inspection of nanosurfaces is essential to ensure the quality of nanostructure manufacturing. Angle-resolved scatterometry provides a non-invasive inspection method that can be used in-line but often suffers from long acquisition times due to dense angular sampling. This paper addresses the data acquisition challenge by proposing an end-to-end compressed learning framework for 5-level vacancy deficiency detection in zinc oxide nanograss using ARS images. The proposed framework integrates a learnable latitude-based sampling layer with a convolutional neural network, allowing sampling and classification to be jointly optimized during training. The sampling layer exploits the physical structure of ARS patterns and learns informative latitudinal regions, which reduces the sampling search space and improves convergence. Evaluation results show that the proposed approach achieves high and stable deficiency-level classification performance under different noise conditions. Using full ARS images, the model achieves 94.2\% accuracy for five-level deficiency classification and 98.6\% accuracy for separating deficient from non-deficient nanosurfaces. The proposed sampling model matches full-image performance while using up to 90\% fewer angular sampling points. Even when sampling points are reduced by 99.7\%, the classification accuracy decreases by less than 10 percentage points. To further improve training with limited data, we also studied a GAN-based augmentation approach and used GAN-generated data for model pretraining. Augmented data resulted in fast convergence within only a few fine-tuning epochs.
\end{abstract}

\begin{IEEEkeywords}
Angle-resolved scatterometry, compressed learning, deficiency detection, end-to-end learning, GAN
\end{IEEEkeywords}



\section{Introduction}

\IEEEPARstart{S}{urface} characterization is the first step in quality control of manufactured nanosurfaces for the detection of deficiencies in these materials. Traditional techniques for surface characterization such as scanning electron microscopy (SEM) require multiple sample preparation steps, and atomic force microscopy (AFM) may alter delicate nanostructures due to tip–sample interaction \cite{echlin2011handbook, magonov1997characterization}. Among available techniques for surface characterization, scatterometry has shown great potential in nanoscale manufacturing, such as semiconductor fabrication \cite{smilde2012evaluation}.

Scatterometry is a non-contact technique for surface characterization \cite{madsen2015fast, alexe2018model}. It is based on the diffraction of light by surfaces. In this technique, a coherent light source with a single wavelength is directed onto the nanosurface, and the scattering pattern reflected by the nanosurface is recorded by a light-detecting sensor.
Scatterometry is widely used in the semiconductor industry for inspection and metrology of manufactured integrated circuits \cite{raymond2005overview, ko2006overlay}. It provides a non-destructive and high-resolution approach for critical dimension measurement and process monitoring, supporting reliable process control \cite{den2017scatterometry}. In addition, scatterometry has been applied to in-situ monitoring of crystal growth processes \cite{heurlin2015situ}.

Different configurations of scatterometry are available for various applications. There are two common configurations: Mueller matrix ellipsometric-based scatterometry and angular scatterometry. These methods are based on manipulating the light source and changing the light detection angle or the laser illumination angle, respectively \cite{niu2001specular, silver2008angle}.
In angular scatterometry, a single image is formed by consecutively scanning the scattered light via mechanical movement of the light source or the light detector.

Angular scatterometry has been investigated for the characterization of periodic and quasi-periodic nanostructures. Previous studies have shown that angle-dependent scattering patterns contain structural information that can be linked to geometric parameters such as height, pitch, and surface roughness through model-based or numerical reconstruction methods \cite{madsen2016imaging, madsen2016scatterometry}. These approaches typically depend on detailed electromagnetic simulations combined with inverse modeling to estimate structural parameters. While such techniques can achieve high accuracy, they often require significant computational resources and assume well-defined parametric surface models, which may not fully describe irregular or damaged nanostructures.

While angular scatterometry techniques have been explored for nanoscale structural characterization, existing approaches primarily rely on dense angular sampling combined with physics-based reconstruction or parametric inverse modeling. These methods are computationally intensive and are not designed for fast, task-specific decision making. To the best of our knowledge, no prior work has investigated task-driven optimization of ARS angular sampling for direct multi-level deficiency classification using a compressed learning framework.

Our study uses a specific configuration of angular scatterometry called angle-resolved scatterometry (ARS), which was introduced by Zimmermann et al. \cite{zimmermann2012process}.
In this configuration, the light source is a coherent laser beam and it is placed at the pole of the sphere. The nanosurface is placed at the center of the sphere and the laser beam is directed onto the nanosurface.
The light detector is mounted on a spherical frame, and it can change its position and scan the whole sphere to form a single image of the scattered light.
The ARS data in Zimmermann's study are used for the characterization of zinc oxide (ZnO) nanograss for the detection of deficiencies; however, no quantitative analysis is provided on the accuracy of the method.
In addition, the main disadvantage of the ARS is the slow recording time of the data because of sequential scanning, which can limit its application in real-world scenarios.

Compressed sensing (CS) and sparse sampling have been widely used to reduce acquisition time and computational cost in many measurement systems. Traditional CS approaches typically rely on random or fixed linear projections to acquire fewer measurements while preserving signal information \cite{candes2006robust, donoho2006compressed}. 
Recent studies on data-driven sampling have shown that learning sampling patterns tailored to a specific downstream task can outperform fixed or random sampling strategies \cite{adler2018learned}.

Considering the above challenges, our study follows two main objectives. The first objective is to accurately classify ARS measurements for deficiency detection in ZnO nanograss. The second objective is to reduce the number of sampling points required to record a single ARS sample while keeping reliable classification performance. Reducing the number of sampling points is important because it directly decreases the acquisition time and improves the practicality of ARS in real-world inspection settings.

To achieve both objectives in a unified framework, we employ compressed learning (CL) \cite{calderbank2009compressed}. Compressed learning is a framework that combines compressed sensing with learning-based methods, where the model learns directly from compressively sampled measurements instead of requiring full-sized data reconstruction. In this approach, the sampling process and the downstream task, such as classification, can be optimized together in a unified framework \cite{adler2016compressed}. These benefits are particularly relevant to ARS, where each measurement angle increases acquisition time.


Compressed learning has been widely studied in several application domains where measurements are expensive or slow to acquire \cite{hollis2018compressed}. For example, it has been applied in imaging problems such as compressive image classification and reconstruction-free recognition \cite{lohit2016direct, zisselman2018compressed}. Compressed learning has also been investigated in biomedical applications, such as electrocardiogram monitoring, where classification is performed directly from compressed measurements to reduce sensing, transmission, and computation cost \cite{li2021enabling}.

In conventional compressed sensing and compressed learning, measurements are modeled as linear projections of a  signal onto a sensing matrix. Such projections are not physically realizable in ARS systems, where measurements are recorded sequentially at individual angular positions. Therefore, the sampling process in our work is formulated as the selection of discrete angular measurements, implemented as an element-wise masking operation rather than a dense linear projection.

Motivated to make deficiency level classification faster through the ARS setup, we propose an end-to-end compressed learning neural network for efficient sampling and detection of deficiencies in the ZnO nanograss. 
In our method, a specific sampling scheme tailored to the scattering pattern of the ARS data is utilized. This sampling scheme is implemented as a sampling layer inside a neural network and trained using the straight-through estimator. In addition to the compressive sampling of the data, the model is designed to detect the deficiency level of the nanograss from the sampled ARS data.
To address the limited size of the available ARS dataset, we further design a conditional WGAN-GP to augment the training data by generating additional synthetic ARS samples conditioned on the deficiency labels.

\begin{figure*}[hbt!]
\centerline{\includegraphics[width=\textwidth]{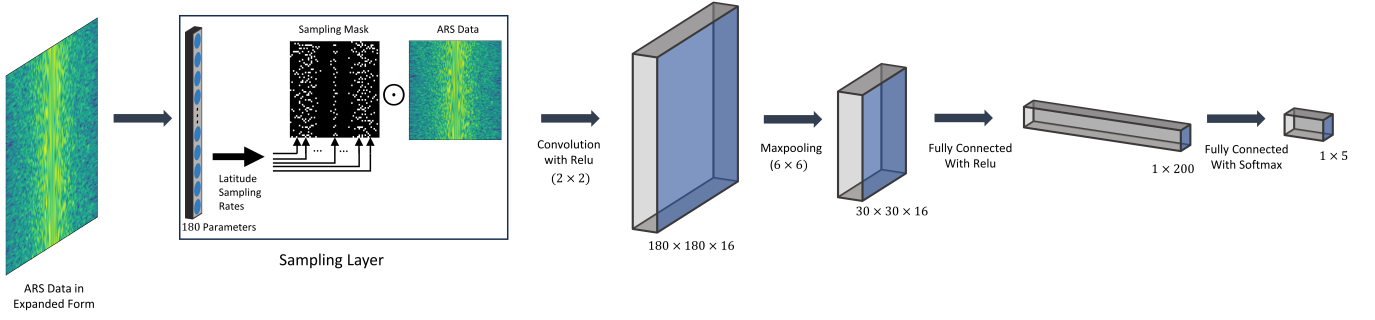}}
\caption{Workflow of the end-to-end compressed learning framework }
\label{fig-workflow}
\end{figure*}

The general overview of the proposed end-to-end CL model is illustrated in Fig.~\ref{fig-workflow}. The model consists of two main parts. The first part includes a layer designed for sampling ARS data. The second part includes a lightweight CNN to perform the deficiency detection task. This framework jointly optimizes the sampling mask to identify the most informative regions in the scatterometry pattern and the classification network within a single end-to-end model.

\noindent The main contributions of this work are summarized as follows:
\begin{itemize}
    \item We propose an end-to-end CL framework for nanograss deficiency detection using ARS images, where classification and sampling strategy optimization are integrated into a single trainable model.



    \item We propose a structured latitude-based sampling scheme that exploits the physical characteristics of ARS scattering patterns. The scheme significantly reduces the sampling optimization search space and provides efficient non-uniform sampling using only $180$ learnable parameters.

    \item We develop a trainable sampling layer based on the proposed scheme and optimize it using a straight-through estimator (STE). This enables gradient-based learning of discrete angular sampling masks despite the non-differentiable sampling operation, allowing fully end-to-end training.

    \item We apply a conditional GAN-based ARS data augmentation approach to address the limited availability of training samples. We further demonstrate that the generated scatterometry images are suitable for pretraining and fine-tuning the proposed classification and sampling model.
\end{itemize}

The remainder of this paper is organized as follows. Section
II explains the proposed end-to-end CL framework and the structure and training process of the GAN  used for data augmentation in detail. 
Section III presents the evaluation of the framework under
various conditions and reports the results. Section IV provides a deeper discussion on the proposed methodology and evaluation results. Finally, Section V summarizes the conclusions of this study.


\section{Methodology}

In this section, the architecture of our proposed model for nanosurface deficiency detection and data  augmentation is discussed in detail. Firstly, the data simulation setup is described to generate the ARS dataset with different levels of deficiencies. Then, the structure of the end-to-end neural network, which is used to simultaneously carry out optimization and classification, is introduced. Next, the straight-through estimator for training the sampling layer in the network is discussed. Lastly, the GAN model and its training process are described in detail.

\subsection{ARS Dataset}

We use ARS data from zinc oxide (ZnO) nanosurfaces with different deficiency levels. This dataset is used to evaluate the accuracy of the proposed method for deficiency detection.
Since creating specific deficiencies on nanosurfaces is difficult \cite{robertson2012spatial}, due to the limited control over the concentration and spatial distribution of these deficiencies at the nanoscale, we use only simulated data in this study. Simulations help in the controlled generation of different deficiency levels and provide a consistent dataset for validating the proposed method.

ARS images of ZnO nanograss with vacancy deficiencies at deficiency levels of 0\%, 10\%, 20\%, 40\%, and 60\% were simulated using the Amsterdam Discrete Dipole Approximation (ADDA) \cite{yurkin2011discrete}, an open-source implementation of the discrete dipole approximation (DDA). The DDA algorithm numerically solves the scattering problem based on Maxwell’s equations for arbitrarily shaped nanostructures. Although the ARS dataset is simulated, the optical configurations and nanosurface parameters strictly follow experimentally realizable setups proposed in \cite{rahman2024scatterometric}, where details of the nanosurface modeling are also provided. The ADDA framework reproduces realistic light–matter interactions in ZnO nanostructures and generates physically consistent scatterometry images suitable for validating the proposed deficiency detection approach.

In our setup, a monochromatic laser beam with a wavelength of 405 nm was directed perpendicularly onto the nanograss. The laser beam had a Gaussian intensity profile with a waist radius of 10~$\mu$m (defined as $1/e^2$ radius). A light detector positioned 60 cm away from the nanograss scanned scattered light over the surrounding sphere. The nanograss material was assumed to be on a transparent substrate that allows scattered light detection in both the upper and lower hemispheres: reflected light in the upper hemisphere and transmitted light in the lower. For this study, we only used reflected scattered light. The light detector scanned the upper hemisphere with an angular resolution of 1\textdegree.

\begin{figure}[hbt!]
\centerline{\includegraphics[width=0.7\columnwidth]{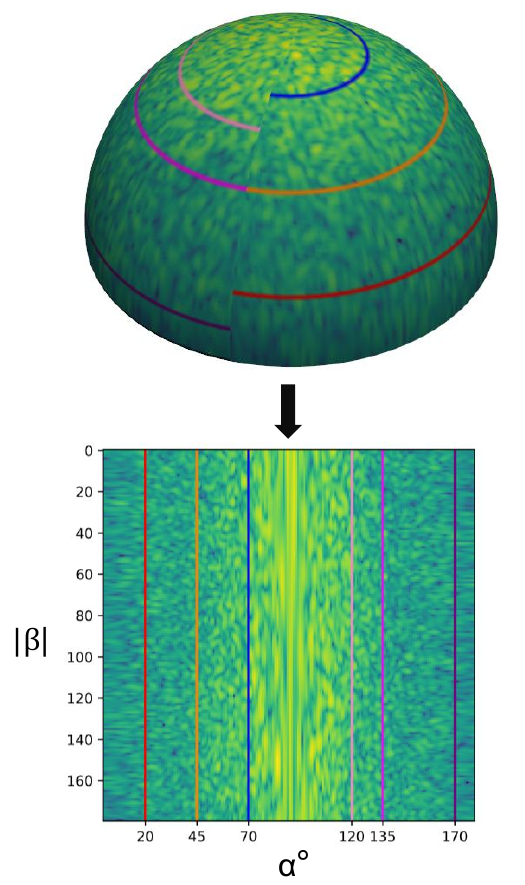}}
\caption{An illustration of ARS data in spherical form and its transformation to the expanded 2D image. A few sample latitudinal arcs are shown in both illustrations.}
\label{fig-ARS}
\end{figure}

A sample image from this dataset is illustrated in Fig.~\ref{fig-ARS} as a hemisphere. For better visualization, an expanded form of the hemisphere is also shown as a 2D image. In this transformation, each latitude of the sphere is represented by a column in the 2D image. The colored arcs in the hemisphere and the corresponding colored columns in the 2D image show identical sampling positions, and they are marked to clearly demonstrate the transformation from the spherical to the 2D representation. In these images, the lighter regions indicate regions where more scattered light from the nanograss is recorded.

We generated a dataset comprising 1400 scatterometry images, with a size of 180 by 180 pixels, that represent five distinct deficiency levels. Specifically, there are 280 images for each deficiency level within this dataset.
These simulated ARS images form the dataset used to train and evaluate the proposed learning model.

\begin{figure}[hbt!]
\centerline{\includegraphics[width=0.8\columnwidth]{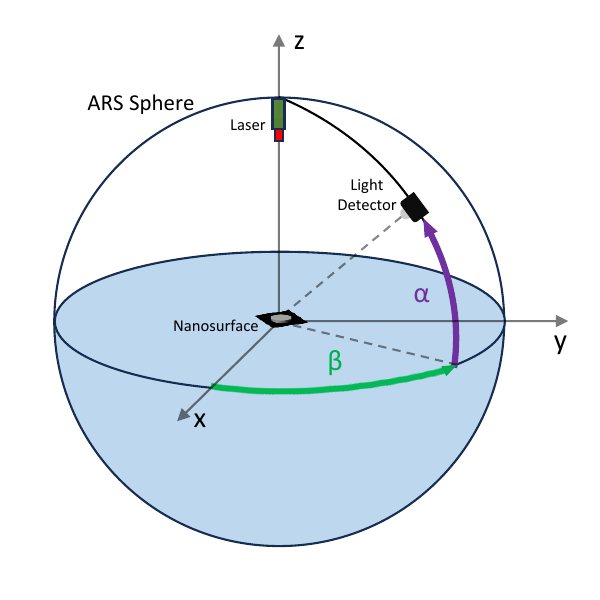}}
\caption{Vertical angle $(\alpha)$ and horizontal angle $(\beta)$ in ARS  sphere }
\label{fig:alpha_beta}
\end{figure}

\subsection{End-to-End Classification and Sampling Model}

Neural networks are well-suited for ARS deficiency detection because scattering patterns are high-dimensional and contain complex, non-linear features that are difficult to model with hand-crafted methods. Therefore, we use a neural network to develop an end-to-end model that jointly optimizes sampling and classification. In our previous work, we relied on hand-crafted features and used a classifier and an optimizer in a sequential loop to iteratively select sampling points and then classify the data \cite{11339492}. In contrast, the proposed end-to-end architecture integrates these two tasks during training. This approach offers two main advantages. First, it is more computationally efficient since sampling and classification are optimized together rather than through separate iterative steps, and second, it carries out automatic feature extraction directly from ARS data, which removes the need to select hand-crafted features.

\subsection{Classification}

The classification part of our model uses a convolutional neural network (CNN) architecture specifically designed for deficiency detection in nanograss using ARS images. 
The CNN consists of a convolutional layer, a pooling layer, and fully connected classification layers.
The model takes single-channel ARS images of size $180 \times 180$ pixels as input and outputs classification probabilities for five deficiency levels ($0\%$, $10\%$, $20\%$, $40\%$, and $60\%$).

The first layer is a 2D convolutional layer with 16 filters of size $2\times2$, using a stride of $1$ and padding of $1$. This configuration allows the network to detect local features in the scattering patterns while maintaining spatial resolution. Following the convolutional layer, a ReLU activation function introduces non-linearity. 
A max-pooling layer with a kernel size of $6\times6$ and a stride of $6$ reduces the spatial dimensions of the feature maps while keeping the most important features. 
The pooled feature maps are then flattened and passed through two fully connected layers. The first fully connected layer outputs 200 features. The second fully connected layer maps these 200 features to the final 5-class output corresponding to the different deficiency levels.

This architecture was empirically selected based on preliminary experiments.
The model is trained using stochastic gradient descent (SGD) optimization with a learning rate of 0.001 and the cross-entropy loss function. The training process runs for $500$ epochs with a batch size of $32$.

\subsection{Sampling}

\subsubsection{Sampling Scheme}

For the neural network to find the optimal sampling mask, it needs to search through a combination of sampling points within an image size of 180 by 180 pixels. This results in a large number of potential sampling masks that can challenge the model's convergence during the training phase. The extensive search space not only slows down training but can also lead to suboptimal solutions.

Therefore, we incorporated domain-specific information regarding the scattering patterns of the ARS images to reduce the search space and optimization parameters. This approach is based on the physical understanding of the angle-resolved scatterometry and the characteristic intensity distributions observed in the data.

As shown in Fig.~\ref{fig-ARS}, a consistent pattern is observed across all ARS images. The area around the pole shows the highest light intensity, or scattered light, because the laser beam is directed at the nanosurface precisely at the pole of the sphere. This causes this area to receive most of the reflected light from the nanosurface.

\begin{figure}[hbt!]
\centerline{\includegraphics[width=\columnwidth]{ 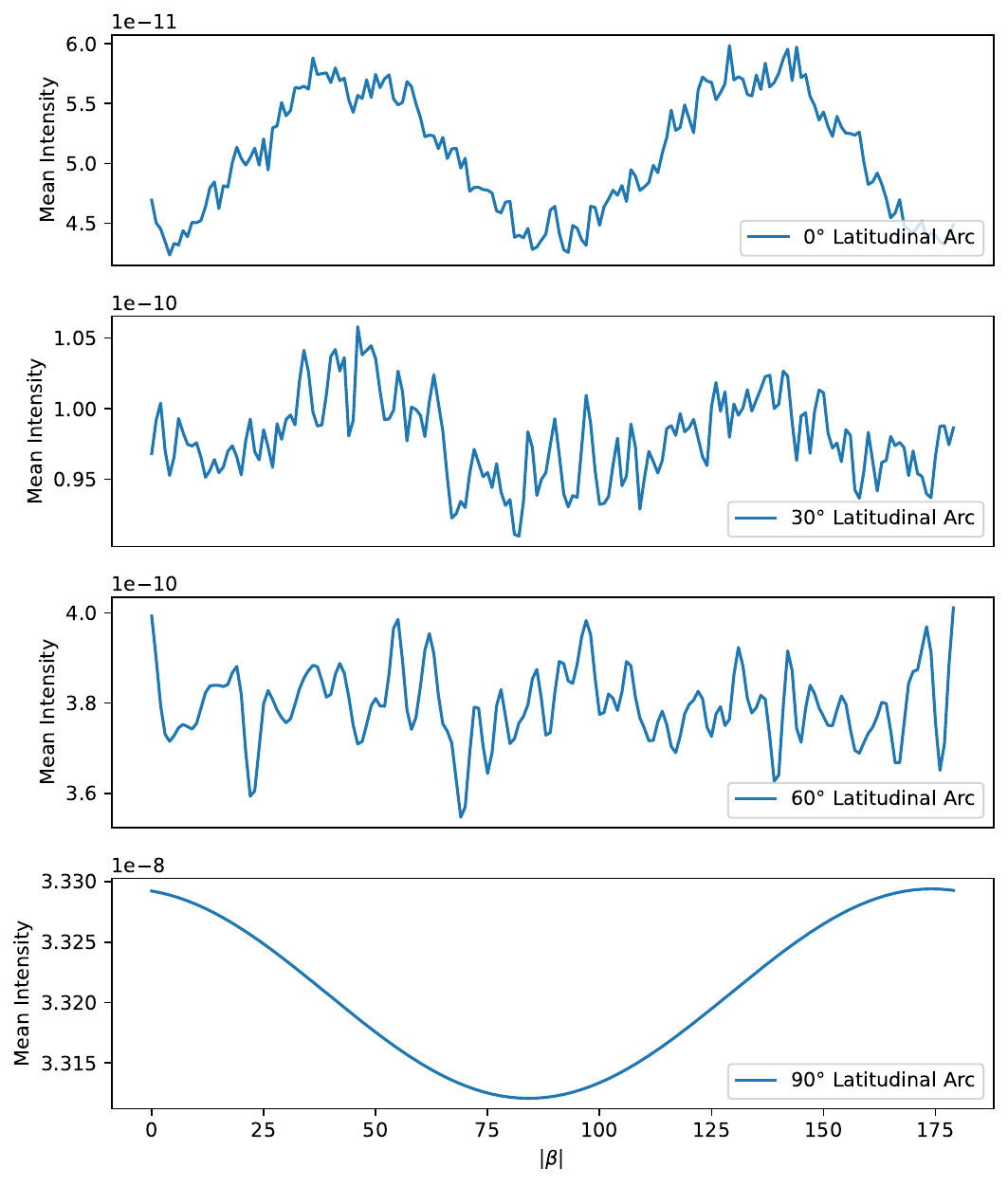 }}
\caption{ Mean intensity distribution across different latitudes in ARS images. The x-axis represents longitudinal arc indices ($|\beta|$), and the y-axis shows the normalized mean intensity values averaged across all longitude positions and multiple sample images.}
\label{fig_latitudes}
\end{figure}

Fig.~\ref{fig_latitudes} shows the mean intensity values for different latitudinal arcs across longitudes. The data from different latitudes exhibit different frequency patterns: polar regions show smoother, low-frequency changes, while equatorial regions show higher-frequency fluctuations. This difference in frequency content means that uniform sampling across the entire surface may not be efficient. Instead, non-uniform sampling can be used to place more sampling points in regions with higher-frequency patterns and fewer points in smoother regions. This approach captures the essential scattering information more effectively.

Based on these observations, we introduce a latitude-based sampling layer that assigns different sampling rates for each latitudinal arc. This approach uses the inherent structure of ARS data, where each latitude (corresponding to a column in the 2D ARS image) represents a consistent angular distance from the pole, allowing the model to learn which angular positions are most informative for classification.

To implement this latitude-based approach, we employ a series of latitudinal arcs that systematically sample the scattering characteristics of the  nanosurface. Instead of using complete latitudinal circles, we divide each latitude into two half-latitudes to increase sampling flexibility. From this point onward, these half-latitudes are referred to as latitudinal arcs, or simply arcs. The first half-latitude spans the horizontal angle $\beta$ from $0^\circ$ to $180^\circ$ at a vertical angle $\alpha = \alpha_1^\circ$, where $\alpha_1$ is selected between $0^\circ$ and $90^\circ$ (see Fig.~\ref{fig:alpha_beta} for angle definitions). The second half-latitude extends through $\beta$ from $0^\circ$ to $-180^\circ$ with a vertical angle $\alpha_2 = 180^\circ - \alpha_1^\circ$. This half-latitude decomposition provides greater sampling flexibility while reducing the total number of required sampling points across the image. Examples of such latitudinal arcs at angles $\alpha = [20^\circ, 45^\circ, 70^\circ, 120^\circ, 135^\circ, 170^\circ]$ are illustrated in Fig.~\ref{fig-ARS}, each shown in distinct colors in both spherical and 2D expanded representations.

We implement this latitude-based sampling strategy using a learnable sampling layer in the end-to-end model. The layer contains $180$ parameters, each corresponding to one latitudinal arc in the ARS image. These parameters represent sampling rates at different angular distances from the pole.
For example, if arcs around a particular latitude contain distinctive scattering patterns related to nanosurface deficiencies, the corresponding parameters increase during training, leading to denser sampling in those regions. Conversely, arcs that contribute little useful information receive lower sampling rates and are sampled less frequently. As a numerical example, if the sampling rate of a particular arc is learned to be $0.8$, approximately $80\%$ of the sampling points along that arc are retained ($144$ out of $180$ positions), whereas an arc with a sampling rate of $0.2$ keeps only about $20\%$ of its points ($36$ out of $180$ positions). In this way, the model automatically focuses on the most informative regions of the ARS image.
These sampling rates are then used to construct a binary mask and are applied to the images in the sampling layer through element-wise multiplication (Hadamard product).

To control the sampling density, we define a maximum sampling budget called maximum samples ($N_{max}$), which sets an upper limit on the total number of sampling points. In the custom sampling function, we normalize the vector of learnable sampling rates so that their sum equals this maximum budget. This normalization prevents the model from exceeding the predefined limit, regardless of how the sampling rates change during training. As a result, $N_{max}$ controls the sampling density and provides a fair comparison between experiments, since all classifiers are trained and evaluated under the same maximum sampling budget across different validation folds and noise conditions.

\subsubsection{Straight Through Estimator}

In our sampling layer, which is integrated into an end-to-end classification model, the continuous sampling rate parameters are converted into binary masks through a series of discrete operations, including spacing calculations, indexing operations, and Hadamard product for pixel selection. These masking operations are essential for implementing the non-uniform sampling strategy but are inherently non-differentiable with respect to the sampling parameters. Traditional backpropagation algorithms require differentiable functions in all layers to compute gradients using the chain rule. However, the presence of these discrete masking operations introduces non-differentiability, which poses a challenge to gradient-based optimization methods.

\begin{figure}[hbt!]
\centering
\includegraphics[width=0.8\columnwidth]{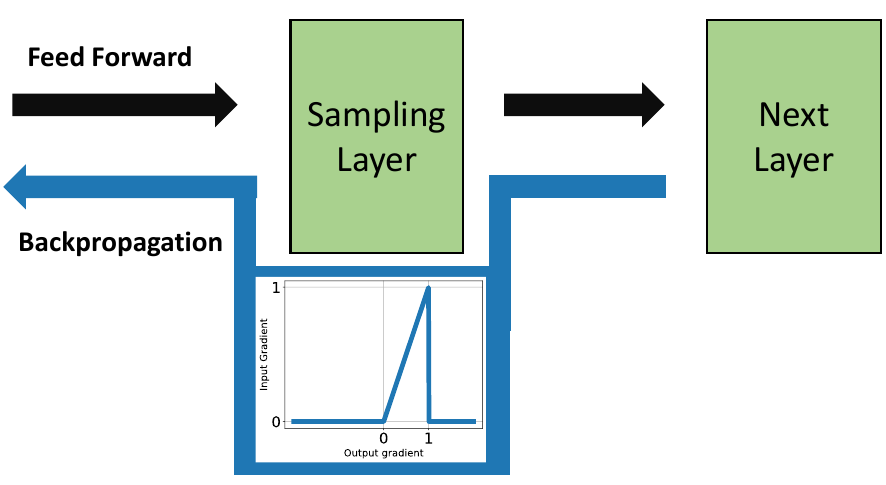}
\caption{Straight-Through Estimator (STE) applied to the sampling layer in a neural network, which allows gradient backpropagation through a non-differentiable operation.}
\label{fig:STE}
\end{figure}

A straight-through estimator (STE) \cite{bengio2013estimating} is a technique commonly used for training neural networks with discrete variables. While the forward pass executes the actual discrete operation, the backward pass approximates the gradient by treating the discrete operation as if it were the identity function. This means that gradients are passed through unchanged, and allows the optimization process to continue despite the presence of non-differentiable components.
As shown in Fig.~\ref{fig:STE}, the STE in our network passes gradients between 0 and 1 without modification during the backward pass, while setting gradients outside this range to zero.


The impact of the STE-based sampling layer is evaluated in Section IV. The results show that the learned sampling patterns focus on regions that contribute most to classification performance while reducing the number of sampled pixels.

\subsection{WGAN-GP-based ARS Data Augmentation}

In recent scatterometry research, the focus has shifted from qualitative interpretation and basic feature-based analysis toward more systematic data processing and machine learning approaches \cite{kolenov2020machine, schmidt2021advanced}. While simple analyses can be performed with a limited number of samples, more detailed studies, such as linking nanosurface defect levels to information extracted from measured ARS data requires larger datasets \cite{hinderhofer2023machine}. This issue becomes more critical when neural networks are used, since their performance strongly depends on the size and variability of the training data \cite{kaplan2020scaling, shorten2019survey}.

In ARS, only a limited number of samples can be derived from the experimental measurements because a certain nanosurface with a certain degree of deficiency is required to record ARS data samples, and prototyping a nanosurface with a specific degree of deficiency is a difficult task \cite{robertson2012spatial}.

ADDA simulations provide an alternative way to generate more samples.
However, this approach is computationally expensive due to the inherent complexity of the Discrete Dipole Approximation. The method discretizes the nanoscale surface into a large number of polarizable dipoles. Then it iteratively solves dense linear systems to accurately model electromagnetic scattering \cite{draine1994discrete}. This process becomes more time-consuming when simulating across multiple angles of incidence, as required in ARS, since each angle needs a separate simulation run \cite{yurkin2007discrete}. Furthermore, the computational complexity increases rapidly with finer spatial resolution and complex nanoscale surface geometries, which require a dense discretization to accurately capture subwavelength features \cite{deprince2010accurate}.

Motivated by the mentioned challenges, we utilize a generative adversarial network (GAN) \cite{goodfellow2020generative} to produce a large dataset to train the proposed end-to-end CL model.
The GAN will be trained on the data, which is simulated with the ADDA algorithm. The objective is to extend the already available but small dataset, which includes scatterometry images of nanosurfaces in 5 deficiency levels, to a bigger one.

A GAN consists of two neural networks: a generator and a discriminator. The generator produces synthetic data samples, while the discriminator tries to distinguish between real and generated samples. Through this adversarial training process, the generator learns to produce outputs that closely match real data \cite{NIPS2014_f033ed80}.

Considering our study's objective, which is to generate data for specific classes or labels, a conditional GAN (cGAN) can be used. In cGANs, both the generator and discriminator take class information as part of their input. As a result, the model generates data conditioned on a selected label \cite{mirza2014conditional}.

\begin{figure*}[hbt!]
\centering
\includegraphics[width=1\linewidth]{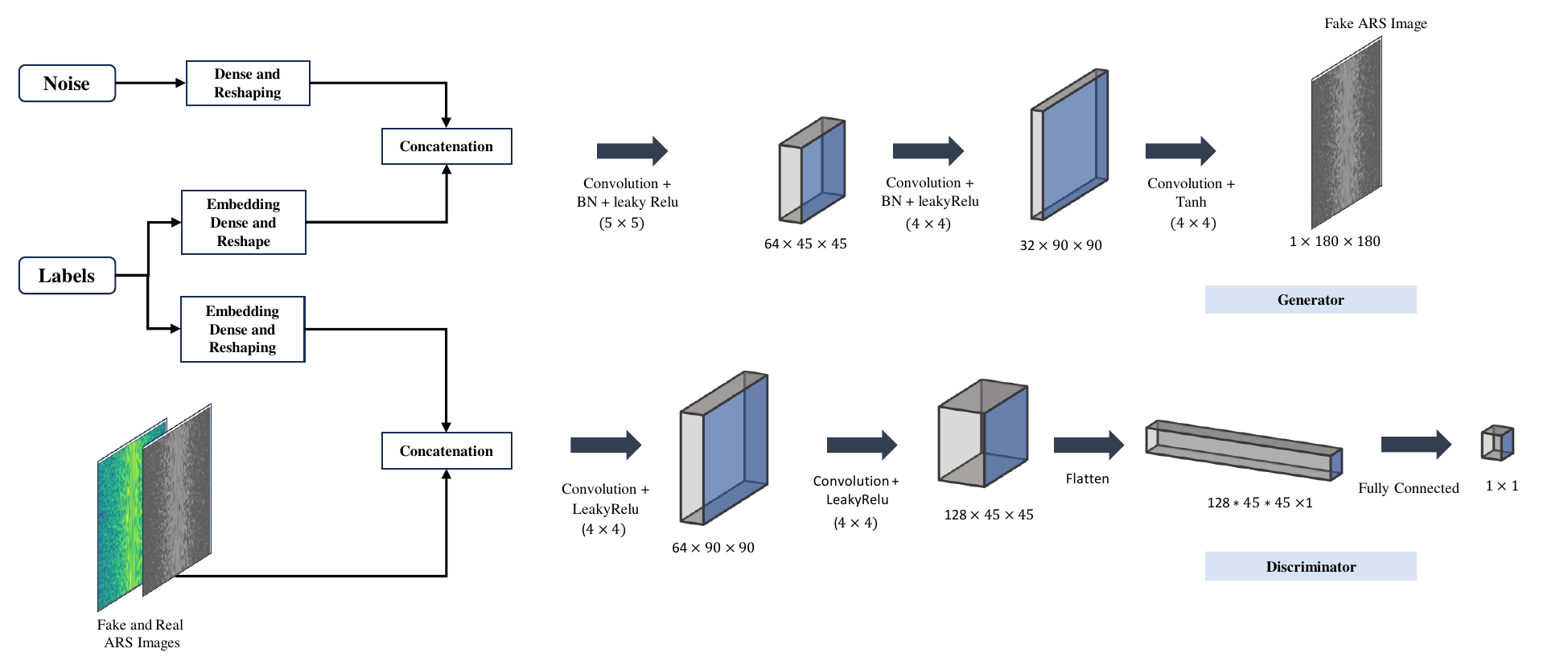}
\caption{Conditional WGAN-GP architecture for generating ARS images corresponding to five nanosurface deficiency levels.}
\label{fig:GAN}
\end{figure*}

Fig.~\ref{fig:GAN} illustrates the architecture of our conditional GAN. The generator takes a random noise vector together with the corresponding class labels to produce synthetic images. The noise input is first preprocessed using a dense layer and reshaped into a feature map. In parallel, the label information is embedded, passed through a dense layer, reshaped, and then concatenated with the noise feature map. The merged representation is then upsampled through convolutional transpose layers to generate the final output image.

The discriminator receives real images and generated fake images, along with the label information. The label embedding is projected and reshaped to match the image size, then concatenated with the image before being fed to convolutional layers. The discriminator outputs a real or fake score, and both networks are optimized through adversarial training.

\subsection{WGAN-GP Training Configuration and Validation Metrics}

\subsubsection{Training Procedure and Loss Function}

In conventional GAN training, the generator and discriminator are optimized simultaneously in a two-player adversarial game. This approach makes the optimization process difficult because the learning target is continuously changing as both networks update their parameters. As a result, standard GAN training often faces two challenges, training instability and mode collapse.

In training instability, the discriminator may become too strong too early, which produces weak gradients for the generator. As a result, the generator cannot learn meaningful patterns, the loss oscillates instead of improving smoothly, and the overall training performance degrades.

In mode collapse, the generator learns a few outputs that can fool the discriminator and repeatedly generates similar samples. As a result, sample diversity is reduced, and the generator fails to represent the full real data distribution.

To overcome the mentioned issues, the proposed GAN's training follows the Wasserstein GAN with gradient penalty (WGAN-GP) framework \cite{gulrajani2017improved, arjovsky2017wassersteingan}. The WGAN-GP objective incorporates a gradient penalty term that enforces the Lipschitz constraint on the discriminator function.

Let $D(\mathbf{x}, y)$ denote the discriminator that outputs a scalar score for an input sample $\mathbf{x}$ conditioned on class label $y$. Here, $\mathbf{x}_r$ denotes real samples drawn from the data distribution $P_r$, while $\mathbf{x}_g$ denotes generated samples following the generator distribution $P_g$. The variable $y$ represents the deficiency class label used for conditioning.

The discriminator loss is formulated as:

\begin{multline}
\mathcal{L}_D =
\mathbb{E}_{\mathbf{x}_g \sim P_g}
\big[D(\mathbf{x}_g, y)\big]
- \mathbb{E}_{\mathbf{x}_r \sim P_r}
\big[D(\mathbf{x}_r, y)\big] \\
+ \lambda_{GP}\mathcal{L}_{GP}
\end{multline}

where $\lambda_{GP} = 10$ denotes the gradient penalty coefficient.

The gradient penalty term is computed as:

\begin{equation}
\mathcal{L}_{GP} = \mathbb{E}_{\hat{\mathbf{x}}}[(\|\nabla_{\hat{\mathbf{x}}} D(\hat{\mathbf{x}}, y)\|_2 - 1)^2]
\end{equation}

where $\hat{\mathbf{x}} = \epsilon \mathbf{x}_r + (1-\epsilon) \mathbf{x}_g$ represents interpolated samples with $\epsilon \sim \text{Uniform}(0,1)$.

The generator loss follows the standard WGAN objective:

\begin{equation}
\mathcal{L}_G = -\mathbb{E}_{\mathbf{x}_g \sim P_g}[D(\mathbf{x}_g, y)]
\end{equation}

To maintain training balance and prevent discriminator overfitting, an asymmetric update schedule is employed with 5 generator updates per discriminator update. The networks are optimized using the Adam optimizer \cite{kingma2014adam} with learning rates of $1 \times 10^{-3}$ for the generator and $5 \times 10^{-6}$ for the discriminator, and beta parameters $\beta_1 = 0.5, \beta_2 = 0.999$.

Training is conducted for 90,000 epochs with a batch size of 64. The latent dimension is set to 100, and the number of deficiency classes is 5, corresponding to the original dataset structure.


\subsubsection{Evaluation Metrics for GAN-Generated Data}

To quantitatively evaluate the quality of the GAN-generated ARS images, multiple validation metrics are used. The mathematical definitions of these metrics are summarized in \mbox{Table~\ref{tab:metrics_formulation}}. Specifically, Fréchet inception distance (FID), maximum mean discrepancy (MMD), and kernel inception distance (KID) measure the distance between the feature distributions of real and generated samples. They provide an indication of how well the generator captures the overall data distribution. In addition, the multi-scale structural similarity index measure (MS-SSIM) is used to evaluate perceptual similarity between generated images, which can show a potential lack of diversity in the generated dataset. These metrics help in the validation of the realism and diversity of the augmented ARS data.

\begin{table*}[htbp]
\centering
\caption{Mathematical formulations of evaluation metrics used for GAN validation.}
\label{tab:metrics_formulation}
\renewcommand{\arraystretch}{1.5}
\begin{tabular}{p{1.1cm} p{5cm} p{10.5cm}}
\toprule
\textbf{Metric} & \textbf{Measures} & \textbf{Mathematical Formulation} \\
\midrule

\multirow{2}{*}{\textbf{FID}} & 
\multirow{2}{5cm}{Distance between feature distributions using Gaussian assumption}
& 
$\text{FID} = \|\boldsymbol{\mu}_r - \boldsymbol{\mu}_g\|_2^2 + \text{Tr}\left(\boldsymbol{\Sigma}_r + \boldsymbol{\Sigma}_g - 2\sqrt{\boldsymbol{\Sigma}_r \boldsymbol{\Sigma}_g}\right)$
\\
& & where $\boldsymbol{\mu}_r, \boldsymbol{\Sigma}_r$ and $\boldsymbol{\mu}_g, \boldsymbol{\Sigma}_g$ are the mean and covariance of features extracted from $\mathbf{x}_r \sim P_r$ and $\mathbf{x}_g \sim P_g$, respectively. \\
\midrule

\multirow{2}{*}{\textbf{MMD}} & 
\multirow{2}{5cm}{Kernel-based distributional distance without parametric assumptions}
& 
$\text{MMD}^2(P_r, P_g) = \mathbb{E}_{x_r,x'_r \sim P_r}[k(x_r,x'_r)] + \mathbb{E}_{x_g,x'_g \sim P_g}[k(x_g,x'_g)] - 2\mathbb{E}_{x_r \sim P_r, x_g \sim P_g}[k(x_r,x_g)]$
\\
& & where $k(\cdot,\cdot)$ is a characteristic kernel function, $P_r$ represents real data distribution, and $P_g$ represents generated data distribution. \\
\midrule

\multirow{2}{*}{\textbf{KID}} & 
\multirow{2}{3.5cm}{Unbiased kernel-based distance using deep features}
& 
$\text{KID} = \text{MMD}^2_u(\mathcal{F}_r, \mathcal{F}_g)$
\\
& & where $\mathcal{F}_r$ and $\mathcal{F}_g$ are features extracted from $\mathbf{x}_r$ and $\mathbf{x}_g$  using Inception-v3, and MMD$_u$ uses an unbiased estimator with polynomial kernel $k(x_1,x_2) = (\gamma x_1^T x_2 + c)^d$. \\
\midrule

\multirow{2}{*}{\textbf{MS-SSIM}} & 
\multirow{2}{5cm}{Multi-scale perceptual similarity between image pairs}
& 
$\text{MS-SSIM}(x_r,x_g) = [l_M(x_r,x_g)]^{\alpha_M} \prod_{j=1}^{M} [c_j(x_r,x_g)]^{\beta_j} [s_j(x_r,x_g)]^{\gamma_j}$
\\
& & where $l_M$, $c_j$, and $s_j$ represent luminance, contrast, and structure comparisons at scale $j$, with weighting exponents $\alpha_M$, $\beta_j$, and $\gamma_j$. $M$ is the number of scales. \\

\bottomrule
\end{tabular}
\end{table*}

\section{Experimental Results}
In this section, the proposed methodology is evaluated through different experiments and from different perspectives to assess its accuracy and robustness. The dataset used for training and testing consists of 1400 data samples. Due to the small size of the dataset used for training, 5-fold cross-validation is used in experiments where needed and relevant to the experiment. We first consider an experiment to confirm the suitability of the classification and sampling methodology to detect the appropriate sampling regions to detect deficiency levels effectively. Then, we evaluate the data generated by the data augmentation framework and study the effect of augmented data on the training of the original sampling and deficiency detection.

\subsection{Deficiency-level Classification Using Full-size ARS Images}
In this experiment, we evaluate the performance of the proposed neural network for only classification without sampling. The objective is to build a baseline for comparison in later stages.

Table~\ref{tab:dataset_noise_results} includes the classification results with different types of noise and different numbers of deficiency levels. To better simulate the real-world scenarios, different levels of noise are introduced to the data. The introduced noises include Gaussian noise with a signal-to-noise ratio (SNR) of $20$~dB and $10$~dB to model the mechanical vibrations in the device and detector sensitivity, and speckle noise with variance of $\sigma^2=0.1$ and $\sigma^2=0.2$ to model multiplicative sensor noise. Additionally, different numbers of distinct deficiency levels were considered by merging specific categories. 
In the $4$-class scenario, the $40\%$ and $60\%$ deficiency levels were merged. 
In the $3$-class scenario, the $40\%$ and $60\%$ levels were merged, as well as the $10\%$ and $20\%$ levels. 
In the $2$-class scenario, the $0\%$ and $10\%$ levels were merged into one class, while the $20\%$, $40\%$, and $60\%$ levels were merged into another class.

Across all class settings, the classification model shows high and stable performance, with accuracy increasing as the number of classes decreases. The 2-class scenario achieves the best and most consistent results, remaining above $98\%$ under all noise conditions, which indicates robustness to both Gaussian and speckle noise. For the 3-class and 4-class tasks, the accuracy drops slightly when noise is added, but the reduction is small and the values remain above $90\%$. The 5-class task is the most affected by noise, with the largest decrease in accuracy under combined Gaussian and speckle noise. However, the performance still stays close to $90\%$, which shows that the model keeps reliable discrimination even under more difficult conditions.

\begin{table*}[ht]
\centering
\caption{Classification accuracy (Avg $\pm$ Std) for ZnO under different noise conditions.  G denotes Gaussian noise and S denotes speckle noise.}
\label{tab:dataset_noise_results}
\begin{tabular}{llccccccc}
\hline
Dataset & Classes & Original & Gaussian 10dB & Gaussian 20dB &  Speckle 0.1 & Speckle 0.2 & G 10dB + S 0.2 & G 20dB + S 0.2 \\
\hline
\multirow{4}{*}{ZnO} 
& 5-class & 94.2$\pm$1.9 & 91.3$\pm$2.4 & 92.5$\pm$2.4 &    91.9$\pm$2.0 & 90.8$\pm$2.6 & 89.2$\pm$2.8 & 90.3$\pm$2.6 \\
& 4-class & 93.8$\pm$1.6 & 90.2$\pm$2.1 & 90.0$\pm$2.2 &    93.8$\pm$1.8 & 91.8$\pm$2.8 & 91.4$\pm$2.0 & 90.6$\pm$2.5 \\
& 3-class & 97.2$\pm$1.3 & 94.6$\pm$1.6 & 94.7$\pm$1.5 &   95.2$\pm$1.5 & 94.5$\pm$1.6 & 94.8$\pm$1.4 & 94.5$\pm$2.0 \\
& 2-class & 98.6$\pm$0.5 & 98.2$\pm$0.3 & 98.2$\pm$0.5 &    98.4$\pm$0.3 & 98.3$\pm$0.4 & 98.2$\pm$0.4 & 98.50$\pm$0.2 \\
\hline
\end{tabular}
\end{table*}

\subsection{Deficiency-level Classification Using Sampled ARS Images}

\begin{figure}[hbt!]
\centerline{\includegraphics[width=.98\columnwidth]{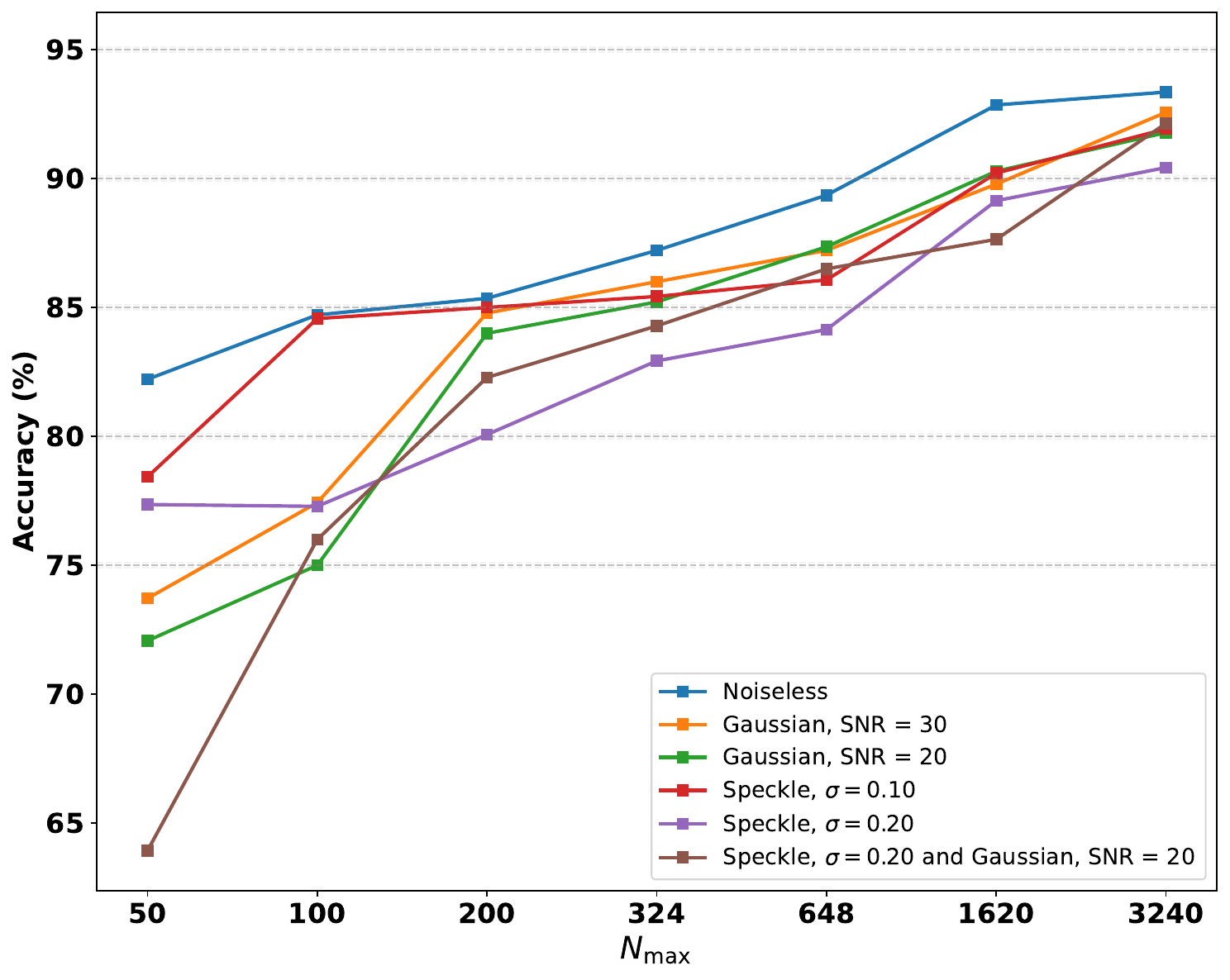}}
\caption{Classification accuracy for different numbers of sampling points under different noise conditions.}
\label{fig:sampling_efficiency}
\end{figure}

Here we evaluate the performance of the end-to-end CL framework in optimizing the sampling positions for a different number of sampling points and effective classification of the deficiency levels.
Fig.~\ref{fig:sampling_efficiency} provides the classification accuracy for different values of $N_{\text{max}}$ (maximum number of sampling points), including $50$, $100$, $200$, $324$, $648$, $1620$, and $3240$ out of a total of $32400$, corresponding to the sampling rates of $0.15\%$, $0.3\%$, $0.6\%$, $1\%$, $2\%$, $5\%$, and $10\%$. Different levels of noise are also considered to study their effect on sampling and deficiency detection accuracy. To acquire each point in the plot, the maximum number of sampling points is set, and the model is evaluated with 5-fold cross-validation. As can be seen in the case of $N_{\text{max}} = 50$, the effect of the different types of noise is more pronounced in the accuracy values.

With an increase in $N_{\text{max}}$ to $100$, a significant increase in the accuracy is observed. As the number of sampling points increases, the accuracy values consistently increase and the plots become closer to each other, meaning that adding more noise does not deteriorate the model's performance compared to a lower number of sampling points.

When comparing the results in Fig.~\ref{fig:sampling_efficiency} with the full-image classification accuracy reported in \mbox{Table~\ref{tab:dataset_noise_results}}, it can be seen that the model with a $10\%$ sampling rate reaches the full-image classification accuracy for both noisy and noiseless data. In addition, in the noiseless case, even a $5\%$ sampling rate achieves the full-image classification accuracy.

\begin{figure*}[hbt!]
\centering
\includegraphics[width=1\linewidth]{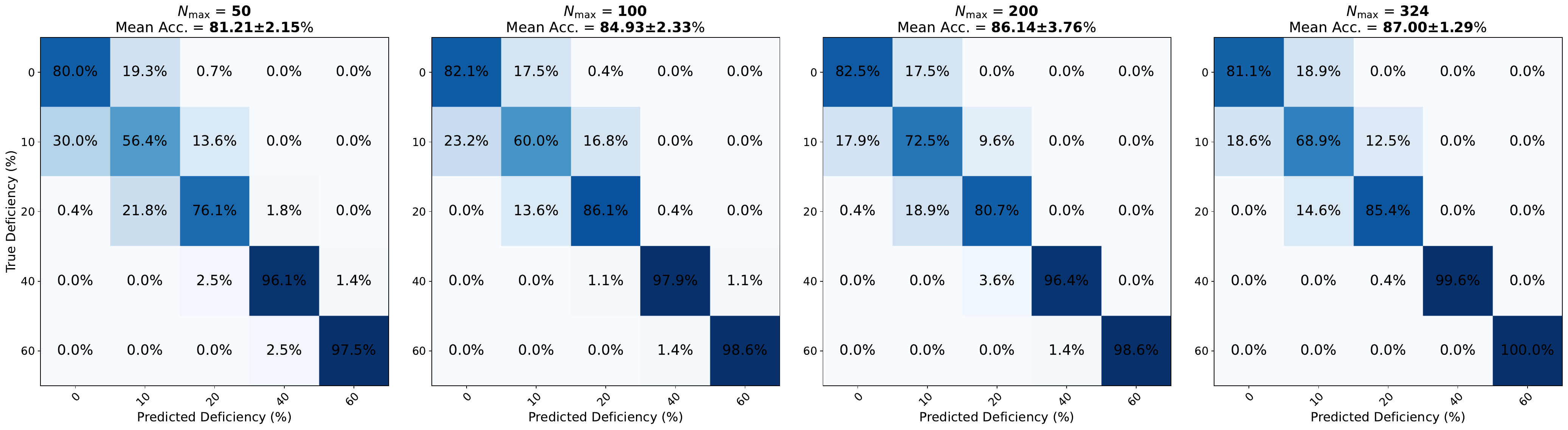}
\caption{Confusion matrices for deficiency level classification on the noiseless dataset using different maximum numbers of sampling points ($50$, $100$, $200$, and $324$).  The mean classification accuracy is reported for each sampling setting on top of each table.}
\label{fig:confusion_matrices}
\end{figure*}

To provide a more detailed view on deficiency level separation results with different $N_{\text{max}}$, the confusion matrices are provided in Fig.~\ref{fig:confusion_matrices} for the noiseless data. The confusion matrix presents the class-wise classification results by comparing the predicted labels with the ground-truth labels. Correct predictions are located on the main diagonal, while off-diagonal entries represent misclassifications between deficiency levels. Each row corresponds to the true class, and each column corresponds to the predicted class.

A key observation across all sampling settings is that the model separates the high deficiency levels ($40\%$ and $60\%$) very well. These classes show that most values are concentrated on the diagonal even at $N_{\text{max}} = 50$ (e.g., $96.1\%$ for $40\%$ and $97.5\%$ for $60\%$), and the performance becomes nearly perfect at $N_{\text{max}} = 324$ ($99.6\%$ for $40\%$ and $100\%$ for $60\%$). This indicates that higher deficiency level signatures on the scatterometry images are highly distinctive and require fewer sampling points for reliable classification.

In contrast, the main classification difficulty is concentrated in the lower deficiency levels, especially between $0\%$ and $10\%$, and between $10\%$ and $20\%$. For example, for the $N_{\text{max}} = 50$ case, the true $10\%$ class is predicted as $0\%$ in $30.0\%$ of cases, and the true $0\%$ class is predicted as $10\%$ in $19.3\%$ of cases. This confusion remains visible even at $N_{\text{max}} = 324$. This suggests that the boundary between non-deficient and mildly deficient levels remains difficult to distinguish.

\subsection{Validation of the GAN-Generated Data}

Validating the GAN-generated scatterometry data is important to verify the reliability and effectiveness of the proposed data augmentation approach. Since the GAN-generated images will be used as training data for the downstream classification model, their quality directly affects the model’s ability to accurately detect deficiency levels in nanosurfaces. Therefore, a comprehensive evaluation framework is needed to measure how well the generated images preserve the statistical properties and structural characteristics of real scatterometry measurements across different deficiency levels.

To assess the quality of GAN-generated scatterometry images, we use multiple metrics that evaluate different aspects of data fidelity. The GAN generates full-resolution $180$ by $180$ pixel scatterometry images.
The evaluation is done by applying different sampling densities to both real and generated full-resolution images, which range from sparse sampling with 50 points to using the complete image data. This approach helps investigate how the quality of the generated images compares to the ADDA-generated ones with different sampling densities.

Table~\ref{tab:results} presents quantitative evaluation results comparing GAN-generated data against ADDA-generated scatterometry images after applying identical sampling procedures to both datasets. For each sampling density, we report four key metrics: FID, MMD, KID, and MS-SSIM. Each cell shows two values: the bold value represents the metric computed between sampled GAN-generated and sampled ADDA-generated images, while the smaller value below provides an ADDA-vs-ADDA baseline computed using the same sampling procedure for reference.

The FID scores reveal a consistent pattern where generative quality decreases as sampling density increases. Our GAN performs best at low sampling densities, achieving FID scores of $0.173$ and $0.181$ for 50 and 100 points, respectively. These values closely match the ADDA-vs-ADDA references of $0.025$ and $0.034$, showing that the GAN captures essential scatterometry features even with sparse sampling. 
FID scores increase with higher sampling densities because, as more points are included, the metric becomes more sensitive to subtle discrepancies between the real and generated feature distributions. However, the GAN maintains performance comparable to ADDA-vs-ADDA baselines across all densities. This demonstrates that the generated data preserves high fidelity across different resolutions.

The MMD results demonstrate consistency across all sampling densities, with values ranging from  $0.571 \times 10^{-3}$ to $0.679 \times 10^{-3}$.  While there is a measurable distributional difference between GAN-generated and ADDA-generated data, this difference remains bounded and small.

KID mean scores show a pattern similar to MMD, with values around $5.5 \times 10^{-3}$ – $6.8 \times 10^{-3}$ across different sampling densities. The ADDA-vs-ADDA KID baselines are close to zero (ranging from $-0.004 \times 10^{-3}$ to $0.008 \times 10^{-3}$), which provides a reference for the lower bound of the metric. Although the ADDA-vs-GAN KID values are not zero, they remain within a small range, indicating only a minor distribution difference between real and generated data. The relatively stable KID performance across sampling densities suggests that the generator preserves important structural features of scatterometry images regardless of the input resolution.

MS-SSIM evaluation was conducted only for full-resolution images, yielding a score of $0.502$ for GAN-generated data compared to the ADDA-vs-ADDA baseline of $0.499$. This performance demonstrates that the generator successfully preserves multi-scale structural information in full-resolution scatterometry images, and captures both fine-grained details and broader spatial patterns, which are important for accurate deficiency classification.

The validation results show that our GAN architecture produces high-quality full-resolution scatterometry data with statistical and structural properties comparable to ADDA-generated data across different sampling scenarios. The generated data also preserves the statistical distribution of the original measurements. These findings confirm that our approach is effective for data augmentation in ARS applications.

\begin{table}[htbp]
\centering
\caption{Quantitative evaluation results across different sampling densities. For each metric, the top value (bold) shows the proposed GAN performance against ADDA-generated data, while the bottom value shows the ADDA-vs-ADDA baseline for comparison.}
\label{tab:results}
\renewcommand{\arraystretch}{1.2} 
\begin{tabular}{c c c c c}
\textbf{Sampling} & \textbf{FID} & \textbf{MMD} & \textbf{KID Mean} & \textbf{MS-SSIM} \\
\textbf{Points} & & ($\times 10^{-3}$) & ($\times 10^{-3}$) & \\
\hline
\multirow{2}{*}{50} & \textbf{0.173} & \textbf{0.679} & \textbf{6.806} & \multirow{2}{*}{--} \\
 & {\scriptsize 0.025} & {\scriptsize 0.079} & {\scriptsize 0.001} & \\
\hline
\multirow{2}{*}{100} & \textbf{0.181} & \textbf{0.600} & \textbf{5.876} & \multirow{2}{*}{--} \\
 & {\scriptsize 0.034} & {\scriptsize 0.076} & {\scriptsize 0.008} & \\
\hline
\multirow{2}{*}{324} & \textbf{0.305} & \textbf{0.611} & \textbf{6.024} & \multirow{2}{*}{--} \\
 & {\scriptsize 0.137} & {\scriptsize 0.073} & {\scriptsize $-$0.004} & \\
\hline
\multirow{2}{*}{648} & \textbf{0.531} & \textbf{0.623} & \textbf{6.148} & \multirow{2}{*}{--} \\
 & {\scriptsize 0.322} & {\scriptsize 0.075} & {\scriptsize $-$0.001} & \\
\hline
\multirow{2}{*}{1620} & \textbf{0.981} & \textbf{0.586} & \textbf{5.627} & \multirow{2}{*}{--} \\
 & {\scriptsize 0.861} & {\scriptsize 0.075} & {\scriptsize $-$0.003} & \\
\hline
\multirow{2}{*}{3240} & \textbf{1.872} & \textbf{0.571} & \textbf{5.469} & \multirow{2}{*}{--} \\
 & {\scriptsize 1.802} & {\scriptsize 0.075} & {\scriptsize $-$0.001} & \\
\hline
\multirow{2}{*}{Full Image} & \textbf{18.655} & \textbf{0.575} & \textbf{5.478} & \textbf{0.502} \\
 & {\scriptsize 18.160} & {\scriptsize 0.074} & {\scriptsize $-$0.001} & {\scriptsize 0.499} \\
\hline
\end{tabular}%
\end{table}

\begin{figure*}[hbt!]
\centering
\includegraphics[width=1\linewidth]{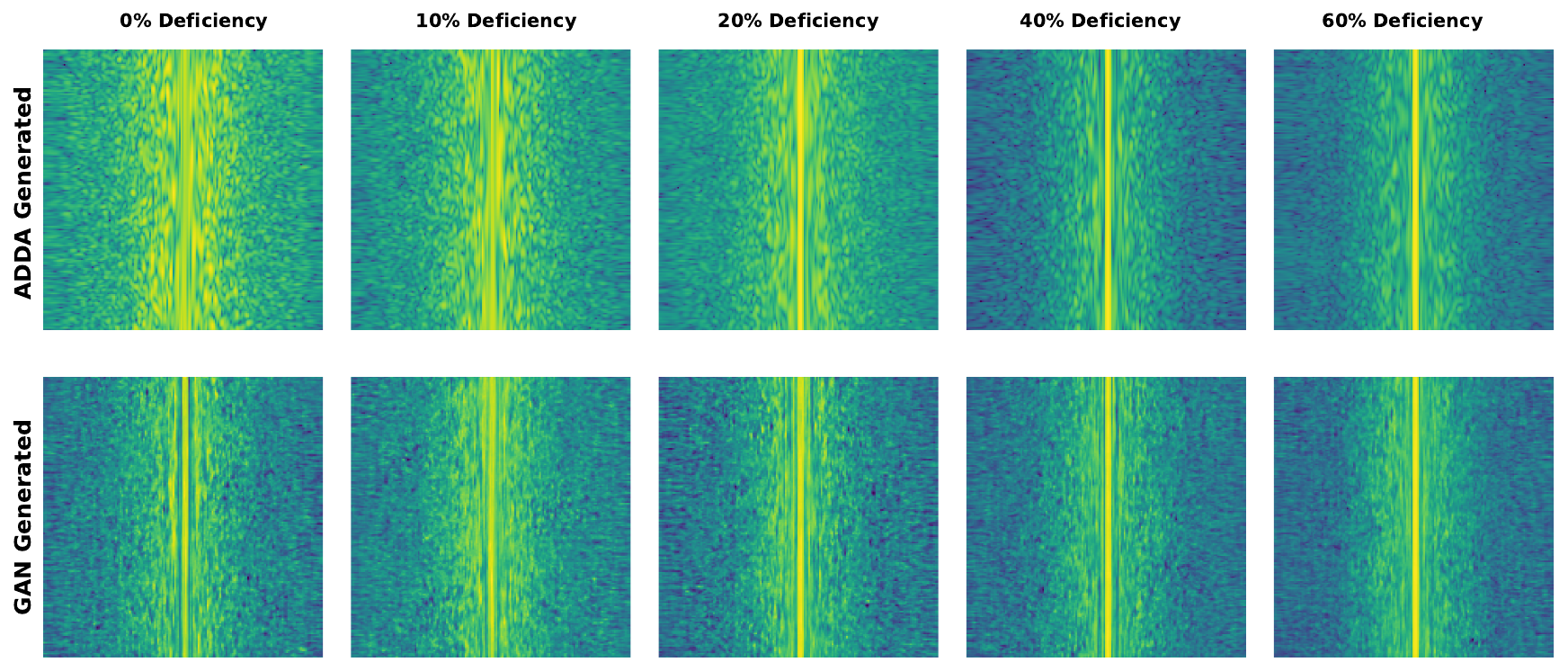}
\caption{Qualitative comparison between ADDA and GAN-generated scatterometry images for the five deficiency classes ($0\%$, $10\%$, $20\%$, $40\%$, and $60\%$). For each class, representative ADDA-generated samples are shown alongside the corresponding synthetic samples.}
\label{fig:GAN_vs_real}
\end{figure*}

To evaluate the visual realism of the GAN-generated scatterometry images, we compare synthetic samples with ADDA-generated samples across the five deficiency levels ($0\%$, $10\%$, $20\%$, $40\%$, and $60\%$) in Fig.~\ref{fig:GAN_vs_real}.
For each class, representative ADDA-generated images and GAN-generated images are shown side by side for visual comparison.

By comparing the ADDA and GAN-generated scatterometry images, it can be seen that the GAN model simulates the general structural pattern of the scattered light well. This includes the symmetric structures and the spatial distribution of the light intensity received by the sensor across different image regions. In particular, both ADDA and GAN generated images show higher intensity in the central area, which is expected because the strongest back-scattered light is concentrated near the region where the laser source is located.

Several class-related features are also well reproduced by the GAN. The overall light intensity decreases as the deficiency level decreases when moving from right to left. In addition, for higher deficiency levels, the scattered light is more concentrated in the central (polar) region, whereas for lower deficiency levels, the light intensity is more widely distributed across different latitudinal arcs in the middle of the images. 

\subsection{ End-to-End CL Model Training with Data Augmentation}

To study the suitability of the augmented data for the classification and sampling model, GAN-generated data is used for pretraining of the proposed deep learning model. In this setup, $6000$ augmented images ($1200$ images per deficiency level) are used to pretrain the model for $5000$ epochs. In the next step, all trainable parameters of the model are set to be frozen, and only the parameters of the two fully connected layers remain trainable for fine-tuning.

\begin{figure*}[hbt!]
\centering
\includegraphics[width=1\linewidth]{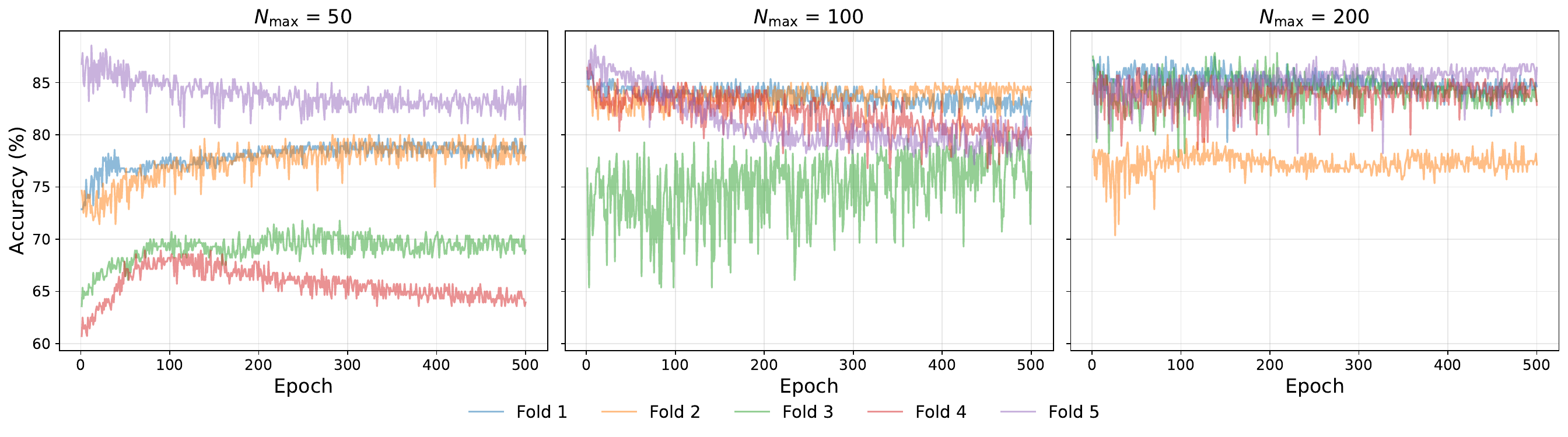}
\caption{Test accuracy over $500$ fine-tuning epochs, evaluated across five folds of the ADDA-generated dataset for different maximum numbers of sampling points.}
\label{fig:GAN_finetuning}
\end{figure*}

For fine-tuning, 1200 ADDA-generated scatterometry images were used for $500$ training epochs. Fig.~\ref{fig:GAN_finetuning} shows the test accuracy over $500$ training epochs for five different folds of the ADDA-generated dataset. It can be seen that, after $500$ epochs, the accuracies obtained for the $N_{\text{max}} = 100$ and $N_{\text{max}} = 200$ cases converge to values that are very close to the accuracies achieved when using only ADDA-generated data, which were $84.93\%$ and $86.76\%$, respectively.

A more interesting outcome of this experiment is the fast convergence to the final accuracy within only a few training epochs. For the $N_{\text{max}} = 100$ and $N_{\text{max}} = 200$ cases, after only $1$ training epoch, mean accuracy values of $83.86 \pm 3.59$ and $84.57 \pm 3.58$ are achieved, respectively. Considering this rapid convergence and that only the final two layers are fine-tuned, these results indicate that the augmented data provides highly informative features for training the first convolutional layer and the sampling layer.

\section{Discussion}

To evaluate the effectiveness of the proposed optimized sampling strategy, random selection of sampling positions was also used. In this approach, at each run, sampling points were selected uniformly at random without incorporating structural information from the data, and this sampling was applied to all data and remained unchanged during training. The test accuracy and standard deviation are reported for different numbers of sampling points in Table~\ref{tab:random_sampling}. Accuracy increases consistently with the number of sampling points, from $61.34\%$ with 50 points to $88.05\%$ with $3240$ points.

The CL model achieves higher accuracy at all sampling levels compared to random sampling. With $50$ sampling points, it reaches $81.21\%$, substantially outperforming random sampling. This advantage is maintained across the full range of sampling densities. The performance gap becomes smaller at higher sampling densities due to redundancy in the data because both methods capture similar information. Nonetheless, for lower numbers of sampling points, which is the main benefit of optimized sampling, a large performance difference is observed.

\begin{table}[h]
\centering
\caption{Test accuracy (mean $\pm$ standard deviation) across 5-fold cross-validation for different numbers of sampling points for random sampling patterns.}
\label{tab:random_sampling}
\setlength{\tabcolsep}{20pt}
\begin{tabular}{c c}
\hline
\textbf{Number of Sampling Points} & \textbf{Test accuracy (\%)} \\
\hline
50   & $61.34 \pm 2.91$ \\
100  & $68.02 \pm 2.10$ \\
200  & $74.55 \pm 1.83$ \\
324  & $78.90 \pm 1.25$ \\
648  & $82.13 \pm 0.98$ \\
1620 & $86.40 \pm 0.71$ \\
3240 & $88.05 \pm 0.52$ \\
\hline
\end{tabular}
\end{table}

\begin{figure}[hbt!]
\centerline{\includegraphics[width=.9\columnwidth]{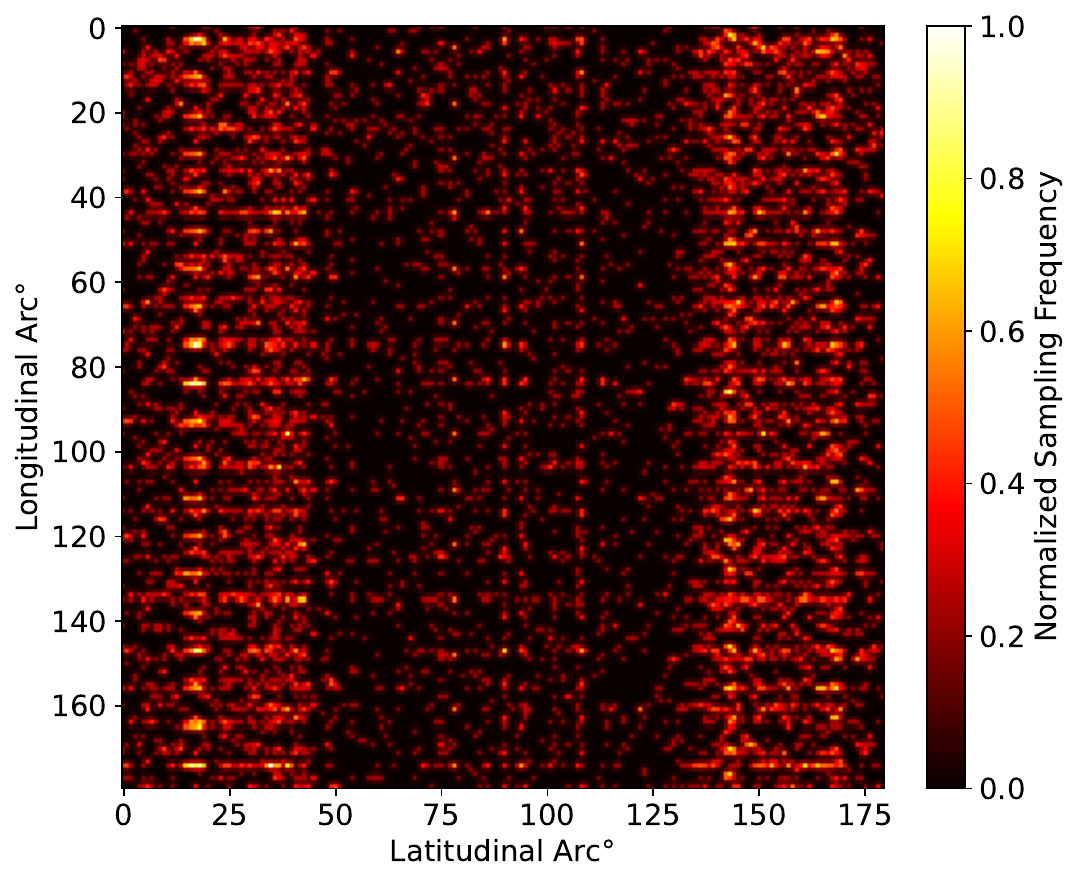}}
\caption{Heatmap of Frequently Sampled Regions }
\label{fig:heatmap}
\end{figure}

Fig.~\ref{fig:heatmap} shows a heatmap formed from the sampled regions over 20 different runs, which displays how often different positions are chosen across the sampling area. The heatmap is displayed as a 2D expanded view of the sphere. In this representation, the center of the image corresponds to the polar regions, while the left and right sides correspond to the equatorial regions, as illustrated in Fig.~\ref{fig-ARS}. Brighter regions indicate more frequently sampled regions, whereas darker regions indicate regions that are sampled less often.

The heatmap shows that the model consistently samples two specific regions. The first corresponds to the polar regions of the ARS sphere, which align with the central columns of the equivalent 2D projected image, where image intensity was highest. The second consists of distinct latitudinal arcs near the equator, specifically within the angular ranges of $15^\circ < \alpha < 45^\circ$ and $135^\circ < \alpha < 165^\circ$ where the scattering pattern was most pronounced and not affected by light speckle superpositions.
The observed pattern exhibits a high degree of symmetry. This symmetrical distribution demonstrates that the algorithm's selection strategy is not biased toward one hemisphere or orientation of the nanosurface.

Table~\ref{tab:dataset_noise_results} shows that the $2$-class scenario achieves almost perfect deficiency detection under all considered noise conditions. This suggests that, when the task is defined as separating non-deficient and deficient nanosurfaces, the ARS patterns provide strong and reliable features even in the presence of Gaussian and speckle noise. This $2$-class setting is also closer to industrial inspection scenarios, where the main goal is to detect whether a nanosurface is deficient, instead of estimating the exact deficiency level.

Fig.~\ref{fig:sampling_efficiency} indicates that the proposed end-to-end CL model can reach full-image classification accuracy while using only a small portion of the ARS image. This suggests that the ARS patterns contain a high level of spatial redundancy and only specific regions include most of the class-discriminative information. Specifically, sampling performance improves fast when moving from extremely sparse sampling toward moderate sampling densities, while additional sampling points provide smaller performance improvements. This shows that the model learns to extract deficiency-relevant features from a limited number of sampling points. This is important for reducing acquisition time and computational cost in practical scatterometry systems.

The learned sampling behavior shown in Fig.~\ref{fig:heatmap} indicates that the sampling layer selects physically meaningful regions of the ARS sphere. The consistent selection of polar regions aligns with the scattering geometry and the expected higher intensity near the laser incidence direction, as also illustrated in Fig.~\ref{fig-ARS}. Furthermore, the selection of distinct latitudinal arcs near the equator suggests that these angular regions contain informative structures for deficiency classification. The symmetry of the learned mask supports that the strategy is not biased toward one hemisphere or nanosurface orientation. Overall, the learned sampling pattern provides interpretability and supports the validity of the sampling design.


The confusion matrices in Fig.~\ref{fig:confusion_matrices} show that most remaining errors are concentrated in the lower deficiency levels. While higher deficiency levels are separated reliably even at sparse sampling, the separation between non-deficient and mildly deficient levels remains challenging. This suggests that mild deficiency levels produce more similar ARS scattering patterns, which limits the separability even when sampling density increases. This result indicates that improving the discrimination of mild deficiencies may require additional measurement information in addition to the reflected ARS patterns, such as a multi-wavelength light source or the  measurements of transmitted scattered light.

Although ADDA simulations provide accurate and reliable modeling of the nanostructures, the evaluation in this work is still limited to simulated data and does not incorporate experimental ARS measurements. Differences between simulated and measured data may occur due to sensor imperfections, calibration errors, mechanical misalignment, and unmodelled variations in the fabricated nanograss structures. Therefore, validating the proposed sampling strategy using real scatterometry measurements is an important direction for future work.  In addition, evaluating the framework on ARS datasets obtained from different nanosurface structures would provide a stronger assessment of generalizability beyond ZnO nanograss. 
For example, the proposed approach could be extended to nanosurface quality and integrity analysis in nanohole arrays characterized by angle-resolved optical measurements \cite{angelini2024angle}, and imaging scatterometry of nanopillar arrays \cite{gawlik2020hyperspectral}. Further validation using multilayer nanostructure optical measurements is also a promising direction supported by advanced simulation and metrological studies \cite{pham2022efficient}.

Finally, future extensions could investigate real-time adaptive sampling strategies for faster inspections, where sampling decisions are updated during acquisition based on partial measurements. Further work may also consider extending the proposed framework toward industrial inspection pipelines and surface quality assessment under realistic production conditions.


\section{Conclusion}
This paper presented an end-to-end deep learning framework for detecting vacancy deficiency levels in ZnO nanograss using ARS data. A CNN-based classifier was combined with a learnable sampling layer to jointly optimize sampling and deficiency classification. To reduce the sampling search space, a latitude-based sampling strategy was introduced based on the physical behavior of scatterometry patterns. The method was evaluated using a simulated dataset with five deficiency levels and different noise settings. Results showed that the classification model is accurate and robust. The sampling model achieved high accuracy with a very small sampling rate, and reached full-image accuracy using only 5\% to 10\% of the data. In addition, WGAN-GP-based data augmentation was validated using multiple metrics and was used for pretraining. The augmented data improved learning speed and helped the model reach final accuracy in only a few epochs. Overall, the results confirm that the proposed method is efficient, accurate, and suitable for deficiency detection in nanosurface scatterometry.

\section{Acknowledgment}

This work was funded by the Deutsche Forschungsgemeinschaft (DFG, German Research Foundation) – 497286574

\bibliography{ref}{}
\bibliographystyle{IEEEtran}

\begin{IEEEbiography}[{\includegraphics[width=1in,height=1.25in,clip,keepaspectratio]{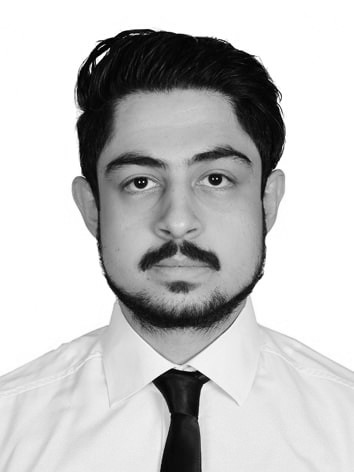}}]{Mehdi Abdollahpour} received the B.Sc. degree in Electrical Engineering (Electronics) and the M.Sc. degree in Biomedical Engineering (Bioelectric) from the University of Tabriz, Iran, in 2017 and 2019, respectively. He is currently pursuing a Ph.D. degree in Electrical Engineering at the University of Bremen, Germany. His research interests include compressed sensing, signal processing, image processing, and machine learning.
\end{IEEEbiography}

\begin{IEEEbiography}[{\includegraphics[width=1in,height=1.25in,clip, trim=6pt 0pt 6pt 0pt, keepaspectratio]{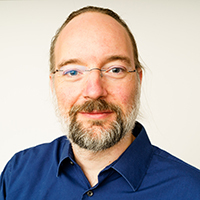}}]{Carsten Bockelmann}  received Dipl.-Ing. and Ph.D. degrees in electrical engineering from the University of Bremen, Germany, in 2006 and 2012, respectively. Since 2012, he has been a Senior Research Group Leader with the University of Bremen, coordinating research activities regarding the application of various frameworks like compressive sensing, DMD, Event-based sampling, and machine learning to communication problems. His research interests include 6G, massive machine-type communication, ultra-reliable low latency communications and industry 4.0 as well as health communications.
\end{IEEEbiography}

\begin{IEEEbiography}[{\includegraphics[width=1in,height=1.25in,clip,trim=0pt 5pt 0pt 0pt,keepaspectratio]{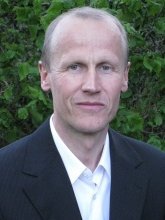}}]{Armin Dekorsy}   (IEEE Senior, 2018) is a professor at the University of Bremen, where he is the director of the Gauss-Olbers Space Technology Transfer Center and heads the Department of Communications Engineering. With over eleven years of industry experience, including distinguished research positions such as DMTS at Bell Labs and Research Coordinator Europe at Qualcomm, he has actively participated in more than 65 international research projects, with leadership roles in 17 of them. He is a Senior Member of the IEEE Communications and Signal Processing Society and a member of the VDE/ITG Expert Committee on Information and System Theory. He co-authored the textbook 'Nachrichtenübertragung, Release 6, Springer Vieweg,' which is a bestseller in the field of communication technologies in German-speaking countries. His research focuses on signal processing and wireless communications for 5G/6G, industrial radio, and 3D networks.
\end{IEEEbiography}


\end{document}